\documentclass[onecolumn,3p]{elsarticle}

\usepackage{graphicx}
\usepackage{amssymb}
\usepackage{amsmath}
\usepackage{hyperref}

\journal{Computer Physics Communications}

\begin{document}

\begin{frontmatter}

\title{Hybrid OpenMP/MPI programs for solving the time-dependent Gross-Pitaevskii equation in a fully anisotropic trap}

\author[ftn]{Bogdan Satari\'{c}\corref{author}}
\ead{bogdan.sataric@uns.ac.rs}

\author[scl-ipb]{Vladimir Slavni\'{c}}
\ead{vladimir.slavnic@ipb.ac.rs}

\author[scl-ipb]{Aleksandar Beli\'{c}}
\ead{aleksandar.belic@ipb.ac.rs}

\author[scl-ipb]{Antun Bala\v{z}}
\ead{antun.balaz@ipb.ac.rs}

\author[bdu]{Paulsamy Muruganandam}
\ead{anand@cnld.bdu.ac.in}

\author[ift]{Sadhan K. Adhikari}
\ead{adhikari@ift.unesp.br}

\cortext[author]{Corresponding author.}
\address[ftn]{Faculty of Technical Sciences, University of Novi Sad, Trg Dositeja Obradovi\'{c}a 6, 21000 Novi Sad, Serbia}
\address[scl-ipb]{Scientific Computing Laboratory, Institute of Physics Belgrade, University of Belgrade, Pregrevica 118, 11080 Belgrade, Serbia}
\address[bdu]{School of Physics, Bharathidasan University, Palkalaiperur Campus, Tiruchirappalli -- 620024, Tamil Nadu, India}
\address[ift]{Instituto de F\'{\i}sica Te\'{o}rica, UNESP -- Universidade Estadual Paulista,  01.140-70 S\~{a}o Paulo, S\~{a}o Paulo, Brazil}

\begin{abstract} 
We present hybrid OpenMP/MPI (Open Multi-Processing/Message Passing Interface) parallelized versions of earlier published C programs (D. Vudragovi\'{c} et al. (2012) \cite{C}) for calculating both stationary and non-stationary solutions of the time-dependent Gross-Pitaevskii (GP) equation in three spatial dimensions.
The GP equation describes the properties of dilute Bose-Einstein condensates at  ultra-cold temperatures.
Hybrid versions of programs use the same algorithms as the C ones, involving real- and imaginary-time propagation based on a split-step Crank-Nicolson method, but consider only a fully-anisotropic three-dimensional GP equation, where algorithmic complexity for large grid sizes necessitates parallelization in order to reduce execution time and/or memory requirements per node. Since distributed memory approach is required to address the latter, we combine MPI programing paradigm with existing OpenMP codes, thus creating fully flexible parallelism within a combined distributed/shared memory model, suitable for different modern computer architectures.
The two presented C/OpenMP/MPI programs for real- and imaginary-time propagation are optimized and accompanied by a customizable makefile.
We present typical scalability results for the provided OpenMP/MPI codes and demonstrate almost linear speedup until inter-process communication time starts to dominate over calculation time per iteration.
Such a scalability study is necessary for large grid sizes in order to determine optimal number of MPI nodes and OpenMP threads per node.
\end{abstract}


\end{frontmatter}

\begin{small}
\noindent
{\bf New version program summary}

\noindent\\
{\em Program title:} GP-SCL-HYB package, consisting of: (i) imagtime3d-hyb, (ii) realtime3d-hyb.\\
{\em Catalogue identifier:} AEDU\_v3\_0     \\
{\em Program Summary URL:} \href{http://cpc.cs.qub.ac.uk/summaries/AEDU_v3_0.html}{http://cpc.cs.qub.ac.uk/summaries/AEDU\_v3\_0.html}\\
{\em Program obtainable from:} CPC Program Library, Queen's University of Belfast, N. Ireland.\\
{\em Licensing provisions:} Apache License 2.0\\
{\em No. of lines in distributed program, including test data, etc.:} 26397.\\
{\em No. of bytes in distributed program, including test data, etc.:} 161195.\\
{\em Distribution format:} tar.gz.\\
{\em Programming language:} C/OpenMP/MPI.\\
{\em Computer:} Any modern computer with C language, OpenMP- and MPI-capable compiler installed.\\
{\em Operating system:} Linux, Unix, Mac OS X, Windows.\\
{\em RAM:} Total memory required to run programs with the supplied input files, distributed over the used MPI nodes: (i) 310 MB, (ii) 400 MB. Larger grid sizes require more memory, which scales with Nx*Ny*Nz.\\
{\em Number of processors used:} No limit, from one to all available CPU cores can used on all MPI nodes.\\
{\em Number of nodes used:} No limit on the number of MPI nodes that can be used. Depending on the grid size of the physical problem and communication overheads, optimal number of MPI nodes and threads per node can be determined by a scalability study for a given hardware platform.\\
{\em Classification:} 2.9, 4.3, 4.12.\\
{\em Catalogue identifier of previous version:} AEDU\_v2\_0.\\
{\em Journal reference of previous version:} Comput. Phys. Commun. 183 (2012) 2021.\\
{\em Does the new version supersede the previous version?:} No.

\noindent\\
{\em Nature of problem:} These programs are designed to solve the 
time-dependent Gross-Pitaevskii (GP) nonlinear partial differential equation in three spatial dimensions in a fully 
anisotropic trap using a hybrid OpenMP/MPI parallelization approach. The GP equation describes the properties 
of a dilute trapped Bose-Einstein condensate.

\noindent\\
{\em Solution method:} The time-dependent GP equation is 
solved by the split-step Crank-Nicolson method using discretization in space 
and time. The discretized equation is then solved by propagation, in 
either imaginary or real time, over small time steps. The method yields 
solutions of stationary and/or non-stationary problems.

\noindent\\
{\em Reasons for the new version:} Previous C \cite{C} and Fortran \cite{Fortran} programs are widely used within the ultracold atoms and nonlinear optics communities, as well as in various other fields \cite{all}. This new version represents extension of the two previously OpenMP-parallelized programs (imagtime3d-th and realtime3d-th) for propagation in imaginary and real time in three spatial dimensions to a hybrid, fully distributed OpenMP/MPI programs (imagtime3d-hyb and realtime3d-hyb). Hybrid extensions of previous OpenMP codes enable interested researchers to numerically study Bose-Einstein condensates in much greater detail (i.e., with much finer resolution) than with OpenMP codes. In OpenMP (threaded) versions of programs, numbers of discretization points in X, Y, and Z directions are bound by the total amount of available memory on a single computing node where the code is being executed. New, hybrid versions of programs are not limited in this way, as large numbers of grid points in each spatial direction can be evenly distributed among the nodes of a cluster, effectively distributing required memory over many MPI nodes. This is the first reason for development of hybrid versions of 3d codes. The second reason for new versions is speedup in the execution of numerical simulations that can be gained by using multiple computing nodes with OpenMP/MPI codes.

\noindent\\
{\em Summary of revisions:}
Two C/OpenMP programs in three spatial dimensions from previous version \cite{C} of the codes (imagtime3d-th and realtime3d-th) are transofrmed and rewritten into a hybrid OpenMP/MPI programs and named imagtime3d-hyb and realtime3d-hyb. The overall structure of two programs is identical. The directory structure of the GP-SCL-HYB package is extended compared to the previous version and now contains a folder scripts, where examples of scripts that can be used to run the programs on a typical MPI cluster are given. The corresponding readme.txt file contains more details. We have also included a makefile with tested and verified settings for most popular MPI compliers, including OpenMPI (Open Message Passing Interface) \cite{OpenMPI} and MPICH (Message Passing Interface Chameleon) \cite{MPICH}.

Transformation from pure OpenMP to a hybrid OpenMP/MPI approach has required that the array containing condensate wavefunction is distributed among MPI nodes of a computer cluster. Several data distribution models have been considered for this purpose, including block distribution and block cyclic distribution of data in a 2d matrix. Finally, we decided to distribute the wavefunction values across different nodes so that each node contains only one slice of the X-dimension data, while containing the complete corresponding Y- and Z-dimension data, as illustrated in Fig.~\ref{fig1}. This allows central functions of our numerical algorithm, calcluy, calcuz, and calcnu to be executed purely in parallel on different MPI nodes of a cluster, without any overhead or communication, as nodes contain all the information for Y- and Z-dimension data in the given X-sub-domain. However, the problem arises when functions calclux, calcrms, and calcmuen need to be executed, as they also operate on the whole X-dimension data. Thus, the need for additional communication arises during the execution of the function calcrms, while in the case of fuctions calclux and calcmuen also the transposition of data between X- and Y-dimensions is necessary, while data in Z dimension have to stay contiguous. Transposition provides nodes with all the necessary X-dimension data to execute functions calclux and calcmuen. However, this needs to be done in each iteration of numerical algorithm, thus necessarily increasing communication overhead of the simulation.

\begin{figure}[!t]
\begin{center}
\includegraphics[width=7.1cm]{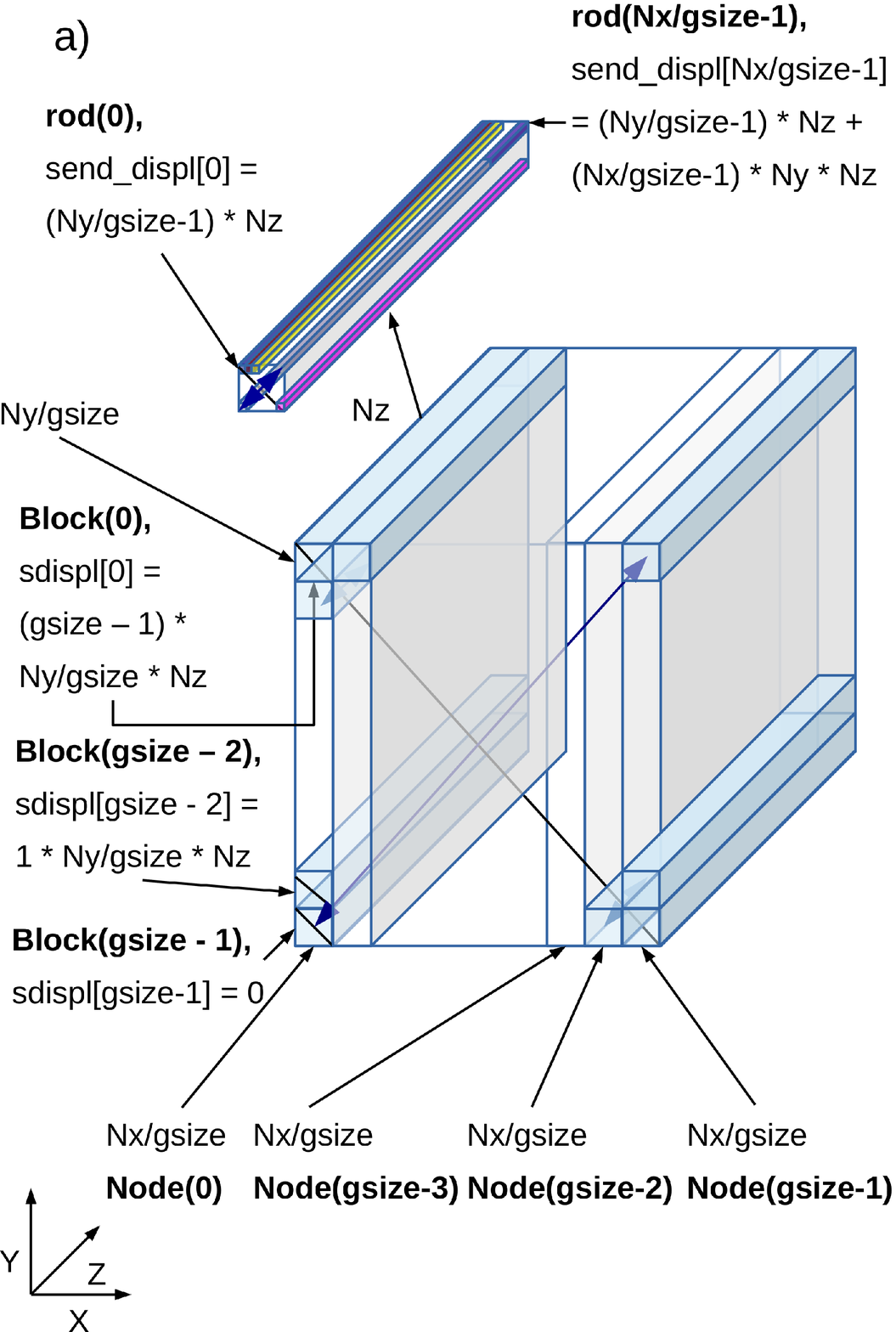}\hspace*{5mm}
\includegraphics[width=7.1cm]{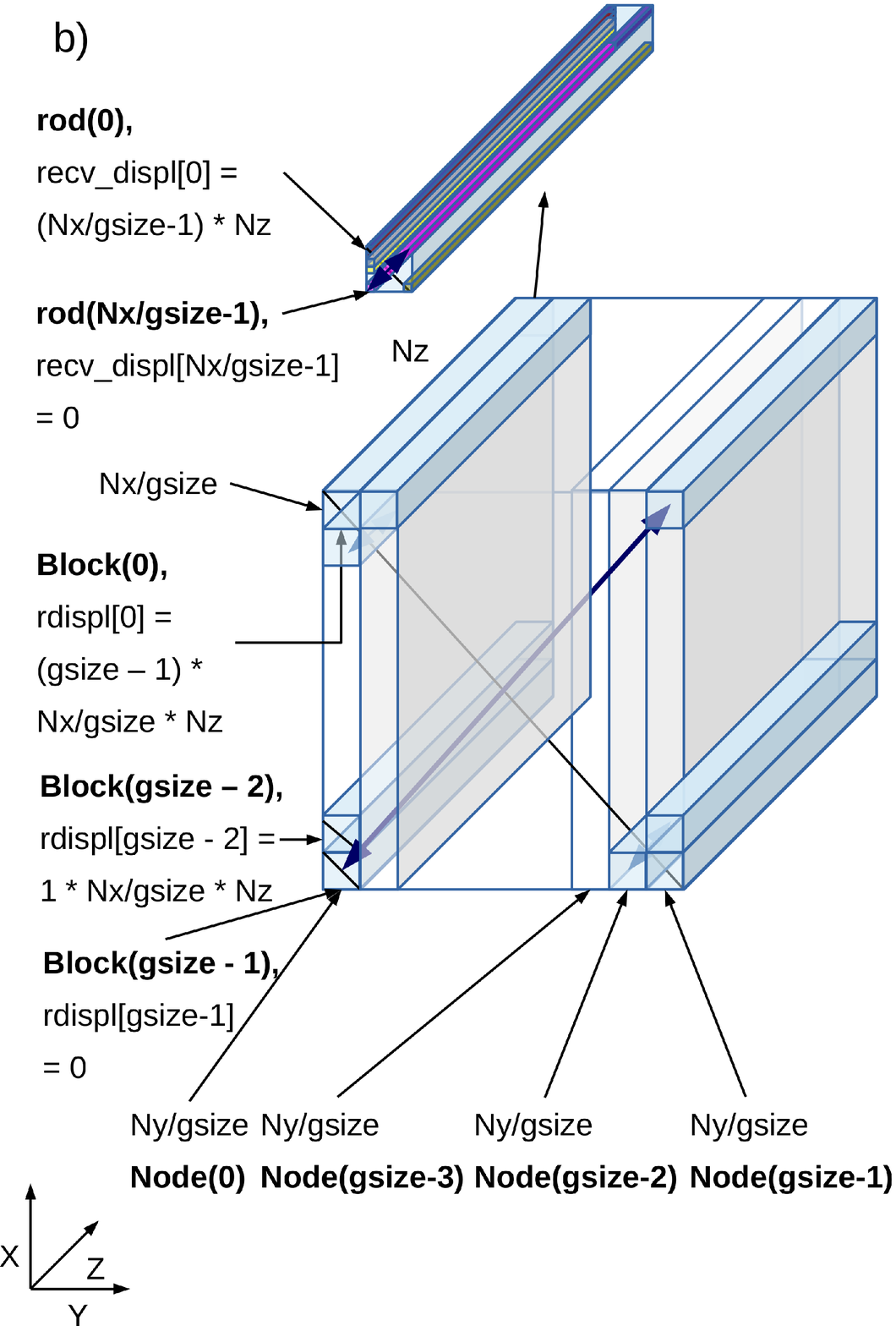}
\caption{BEC wavefunction data structure in the employed algorithm. Data are sliced so that the complete Y- and Z-dimension data reside on a single node for a given range of data in X-dimension, while data in X-dimension are distributed over N = gsize nodes. Figures show data transposition and MPI indexed datatype creation parameters for the case of: (a) sending side and (b) receiving side.}
\label{fig1}
\end{center}
\end{figure}
\begin{figure}[!ht]
\begin{center}
\includegraphics[width=9.3cm]{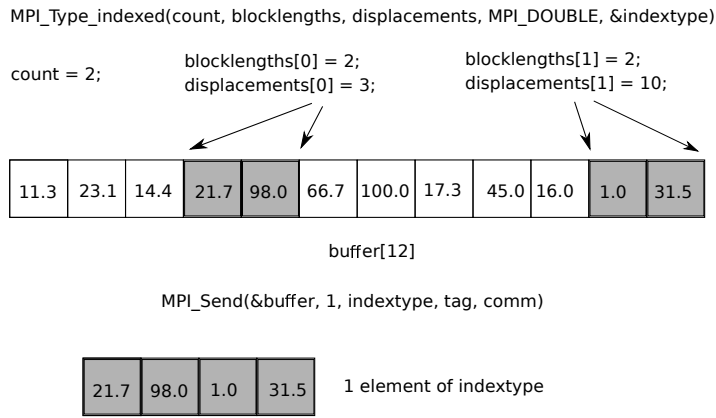}
\caption{Creation of a user-defined MPI datatype indextype with the function MPI\_Type\_indexed. Here, count represents the number of blocks, blocklengths array contains lengths of each block, and displacements array contains the displacement of each block from the beginning of the corresponding data structure. For example, if an array of double precision numbers (designated as buffer in the figure) is sent by MPI\_Send with the datatype set to indextype, it is interpreted as a block-distributed data structure, as specified when indextype was created.}
\label{fig2}
\end{center}
\end{figure}

Transposition algorithms that were considered where the ones that account for greatest common divisor (GCD) between number of nodes in columns (designated by N) and rows (designated by M) of a cluster configured as 2d mash of nodes \cite{dongarra}. Two of such algorithms have been tested and tried for implementation: the case when GCD = 1 and the case when GCD \textgreater 1. The trivial situation N = M = 1 is already covered by the previous, purely OpenMP programs, and therefore, without any loss of generality, we have considered only configurations with number of nodes in X-dimension satisfying N \textgreater 1. Only the former algorithm (GCD = 1) was found to be sound in case where data matrix is not a 2d, but a 3d structure. Latter case was found to be too demanding implementation-wise, since MPI functions and data-types are bound to certain limitations. Therefore, the algorithm with M = 1 nodes in Y-dimension was implemented, as depicted by the wavefunction data structure in Fig.~\ref{fig1}.

\begin{figure}[!b]
\begin{center}
\includegraphics[width=5.6cm]{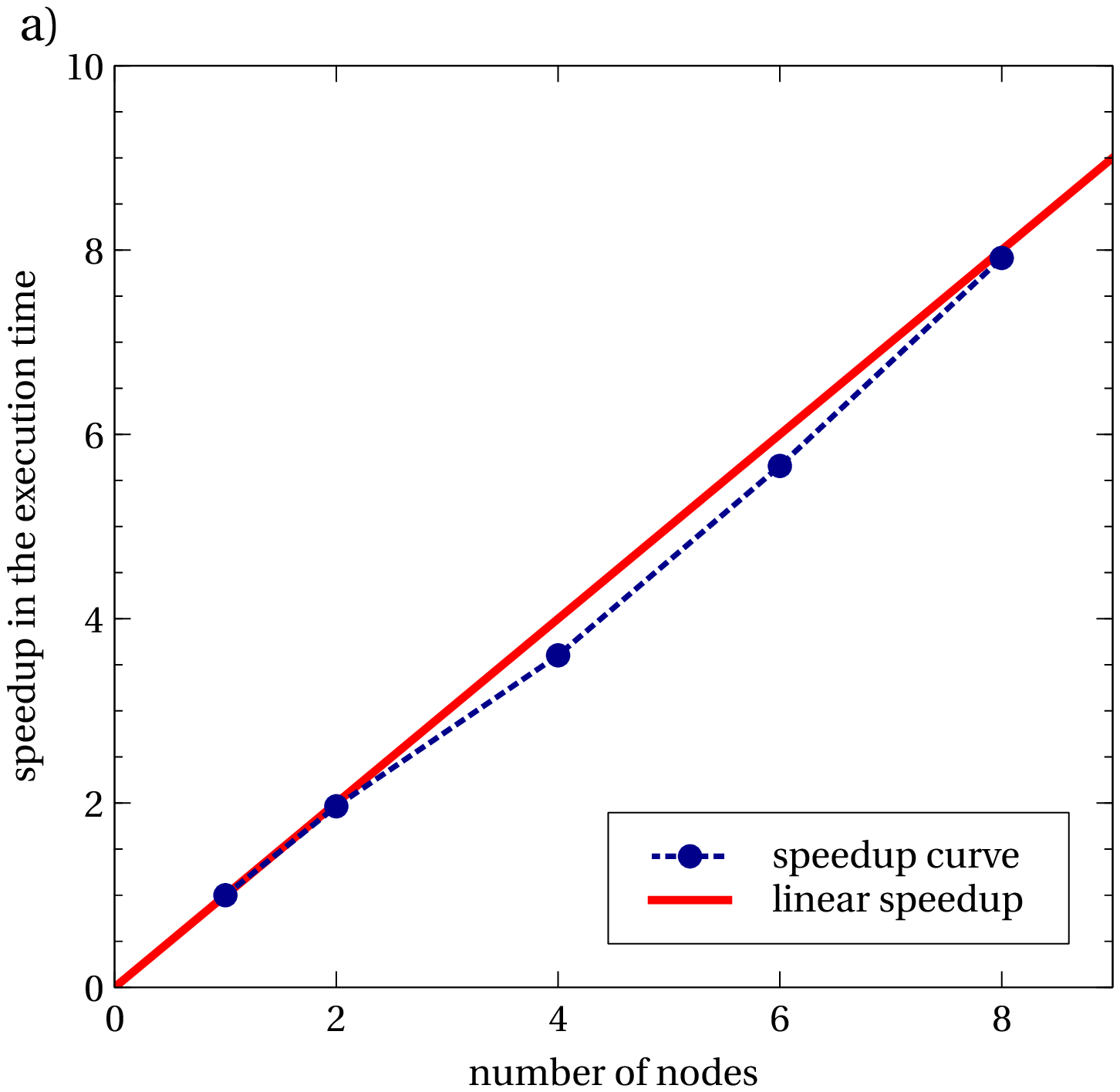}\hspace*{5mm}
\includegraphics[width=5.6cm]{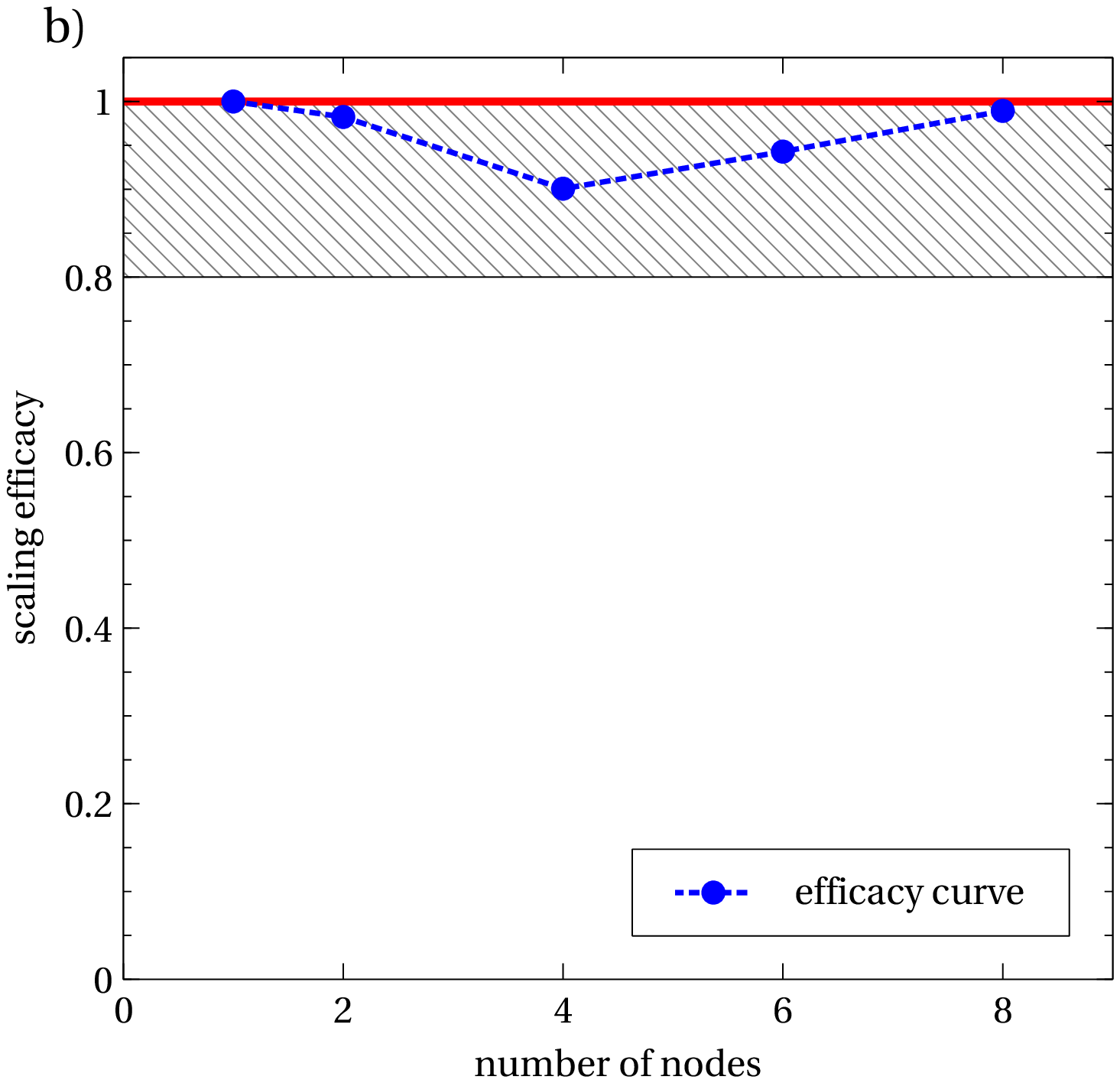}
\includegraphics[width=5.6cm]{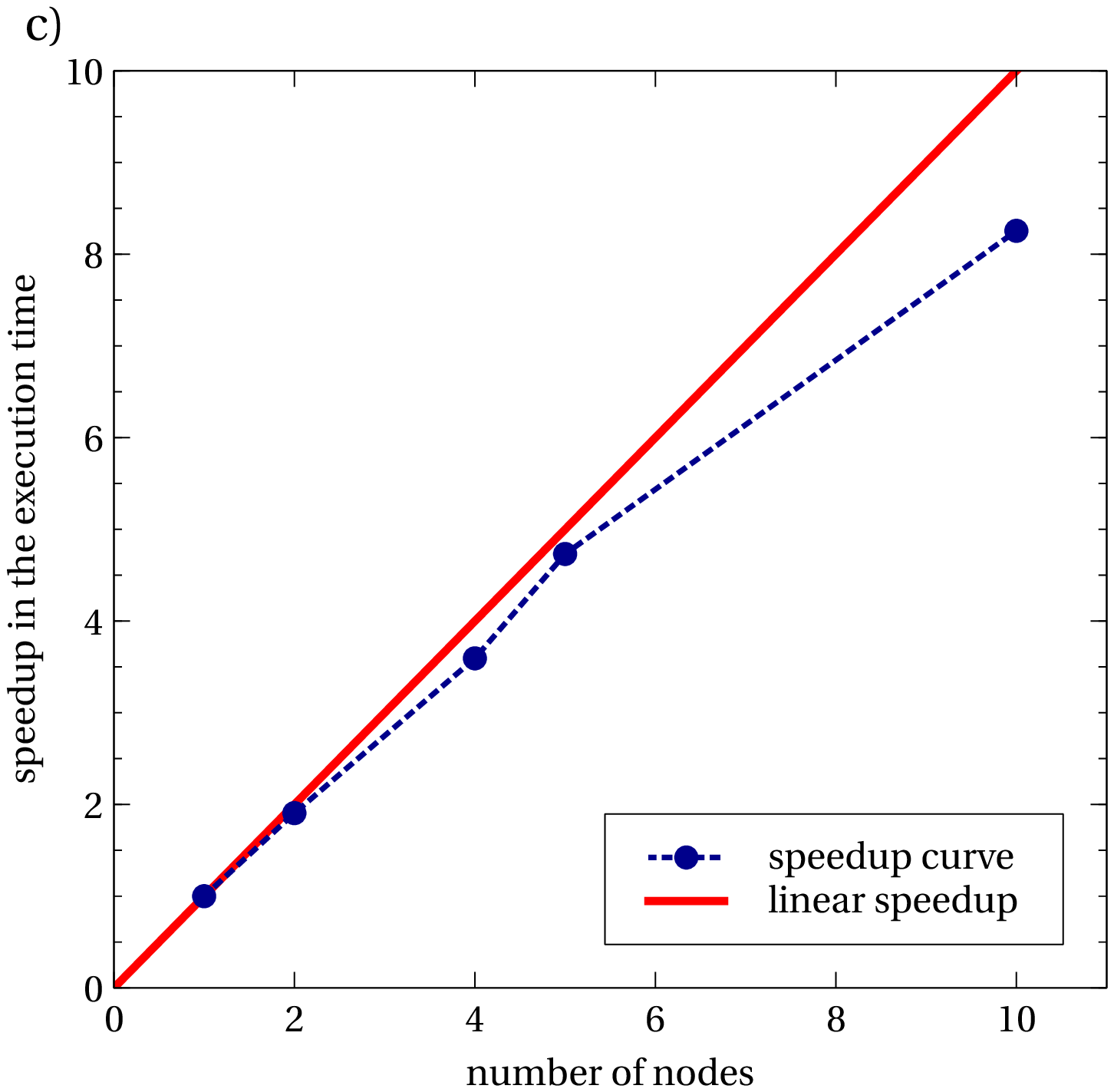}\hspace*{5mm}
\includegraphics[width=5.6cm]{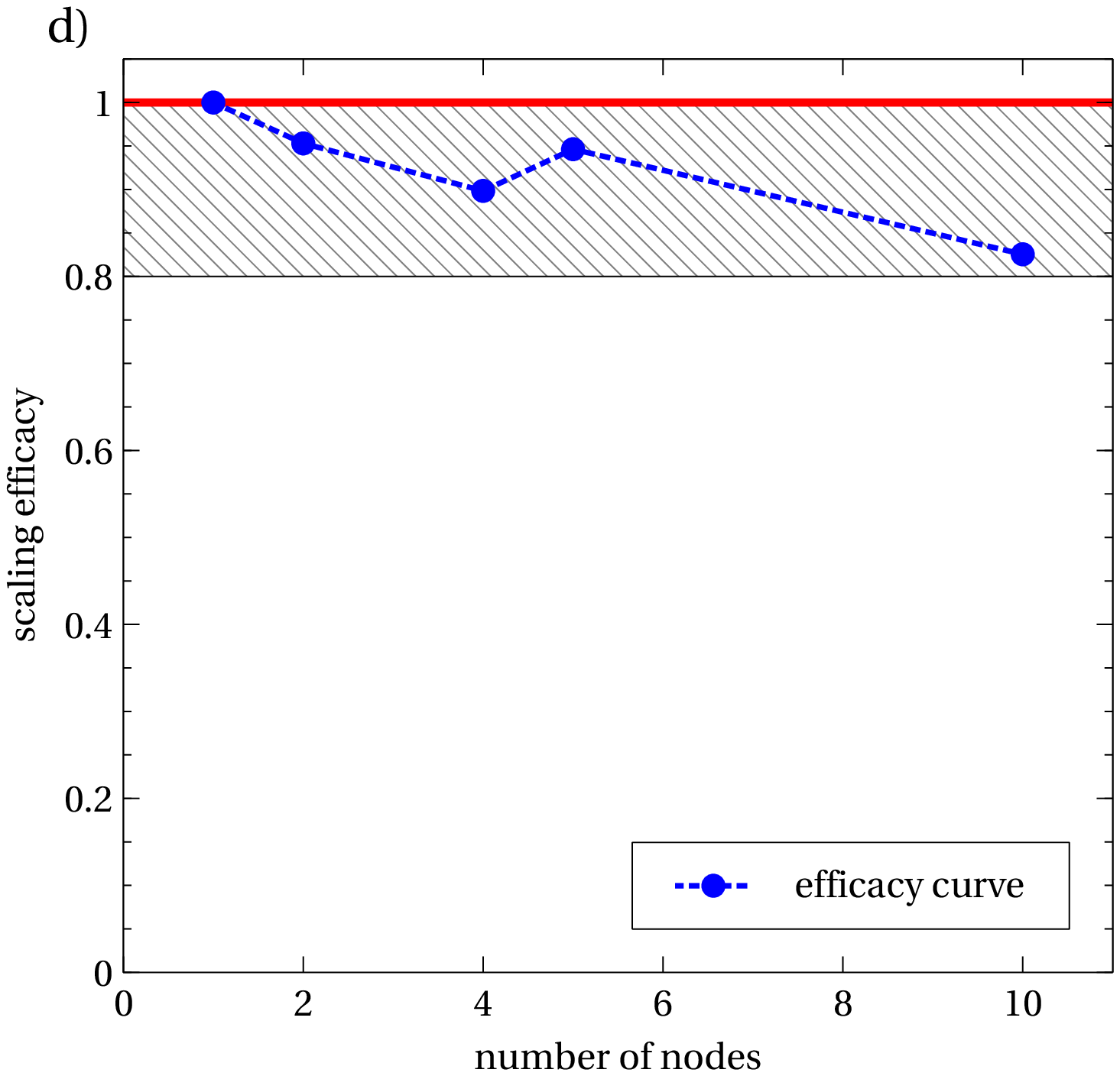}
\caption{Speedup in the execution time and efficacy curves of imagtime3d-hyb and realtime3d-hyb programs as a function of the number of MPI nodes used for small grid sizes. The results are obtained on a cluster with nodes containing 2 x 8-core Sandy Bridge Xeon 2.6 GHz processors with 32 GB of RAM and Infiniband QDR interconnect: (a) speedup of imagtime3d-hyb on a 240x200x160 grid; (b) efficacy of imagtime3d-hyb on a 240x200x160 grid; (c) speedup of realtime3d-hyb on a 200x160x120 grid; (d) efficacy of realtime3d-hyb on a 200x160x120 grid. Shaded areas in graphs (b) and (d) represent high-efficacy regions, where speedup is at least 80\% of the ideal one.}
\label{fig3}
\end{center}
\end{figure}

Implementation of the algorithm relies on a sliced distribution of data among the nodes, as explained in Fig.~\ref{fig2}. This successfully solves the problem of large RAM consumption of 3d codes, which arises even for moderate grid sizes. However, it does not solve the question of data transposition between the nodes. In order to implement the most effective (GCD = 1) transposition algorithm according to Ref.~\cite{dongarra}, we had to carry out block distribution of data within one data slice contained on a single node. This block distribution of data was done implicitly, i.e., data on one node have been put in a single 1d array (psi) of contiguous memory, in which Z-dimension has stride 1, Y-dimension has stride Nz, and X-dimension has stride Ny*Nz. This is different from previous implementation of the programs, where the wavefunction was represented by an explicit 3d array. This change was also introduced in order to more easily form user MPI datatypes, which allow for implicit block distribution of data, and represent 3d blocks of data within 1d data array. These blocks are then swapped between nodes, effectively performing the transposition in X-Y and Y-X directions.

Together with transposition of blocks between the nodes, the block data also have to be redistributed. To illustrate how this works, let us consider example shown in Fig.~\ref{fig1}(a), where one data block has size (Nx/gisze)*(Ny/gsize)*Nz. It represents one 3d data block, swapped between two nodes of a cluster (through one non-blocking MPI\_Isend and one MPI\_Ireceive operation), containing (Nx/gsize)*(Ny/gsize) 1d rods of contiguous Nz data. These rods themselves need to be transposed within the transposed block as well. This means that two levels of transpositions need to be performed. At a single block level, rods have to be transposed (as indicated in upper left corner of Fig.~\ref{fig1}(a) for sending index type and in Fig.~\ref{fig1}(b) for receiving index type). Second level is transposition of blocks between different nodes, which is depicted by blue arrows connecting different blocks in Fig.~\ref{fig1}.

The above described transposition is applied whenever needed in the functions calclux and calcmuen, which require calculations to be done on the whole range of data in X-dimension. When performing renormalization of the wavefunction or calculation of its norm, root-mean-square radius, chemical potential, and energy, collective operations MPI\_Gather and MPI\_Bcast are also used.

\begin{figure}[!b]
\begin{center}
\includegraphics[width=5.6cm]{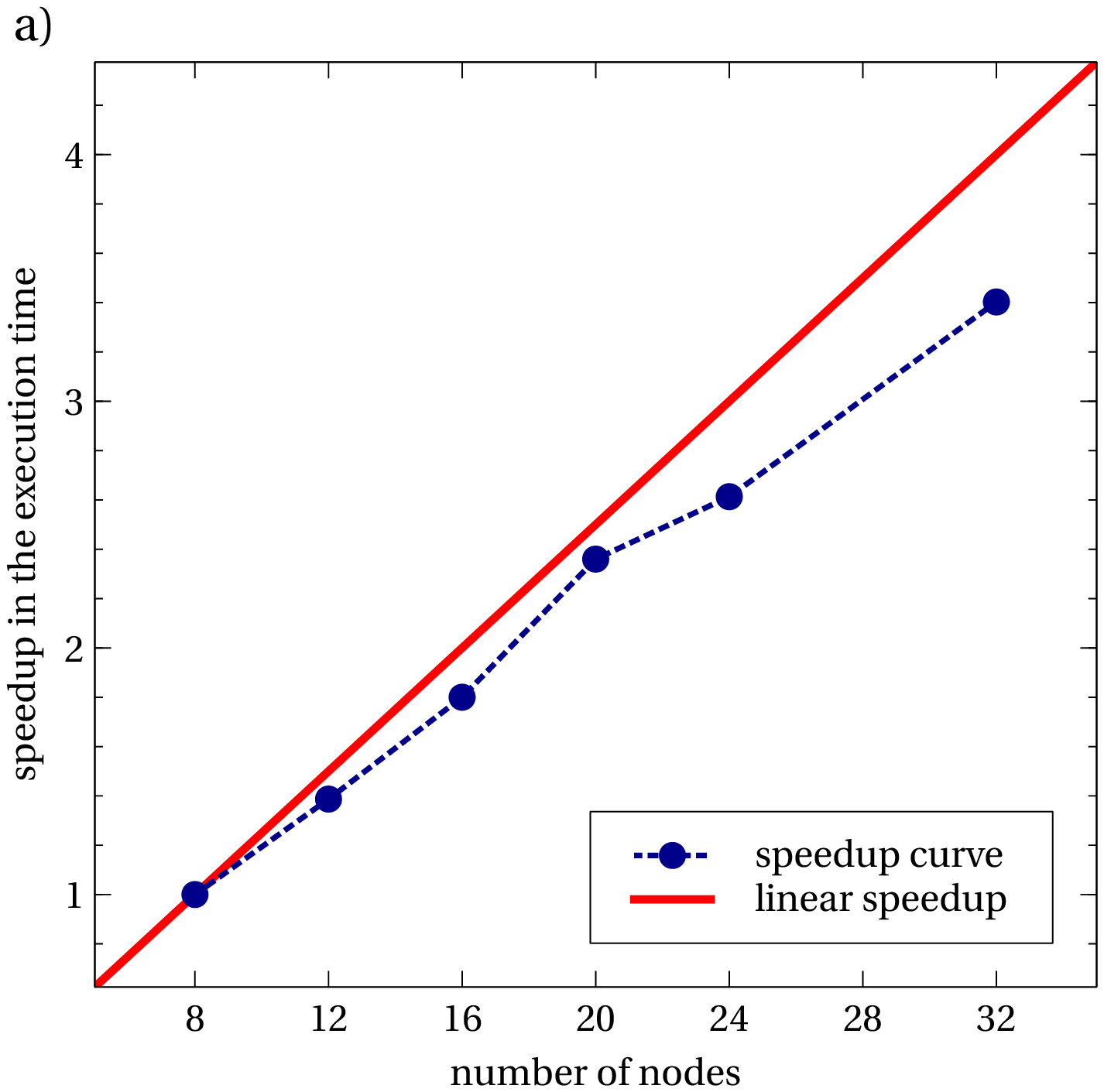}\hspace*{5mm}
\includegraphics[width=5.6cm]{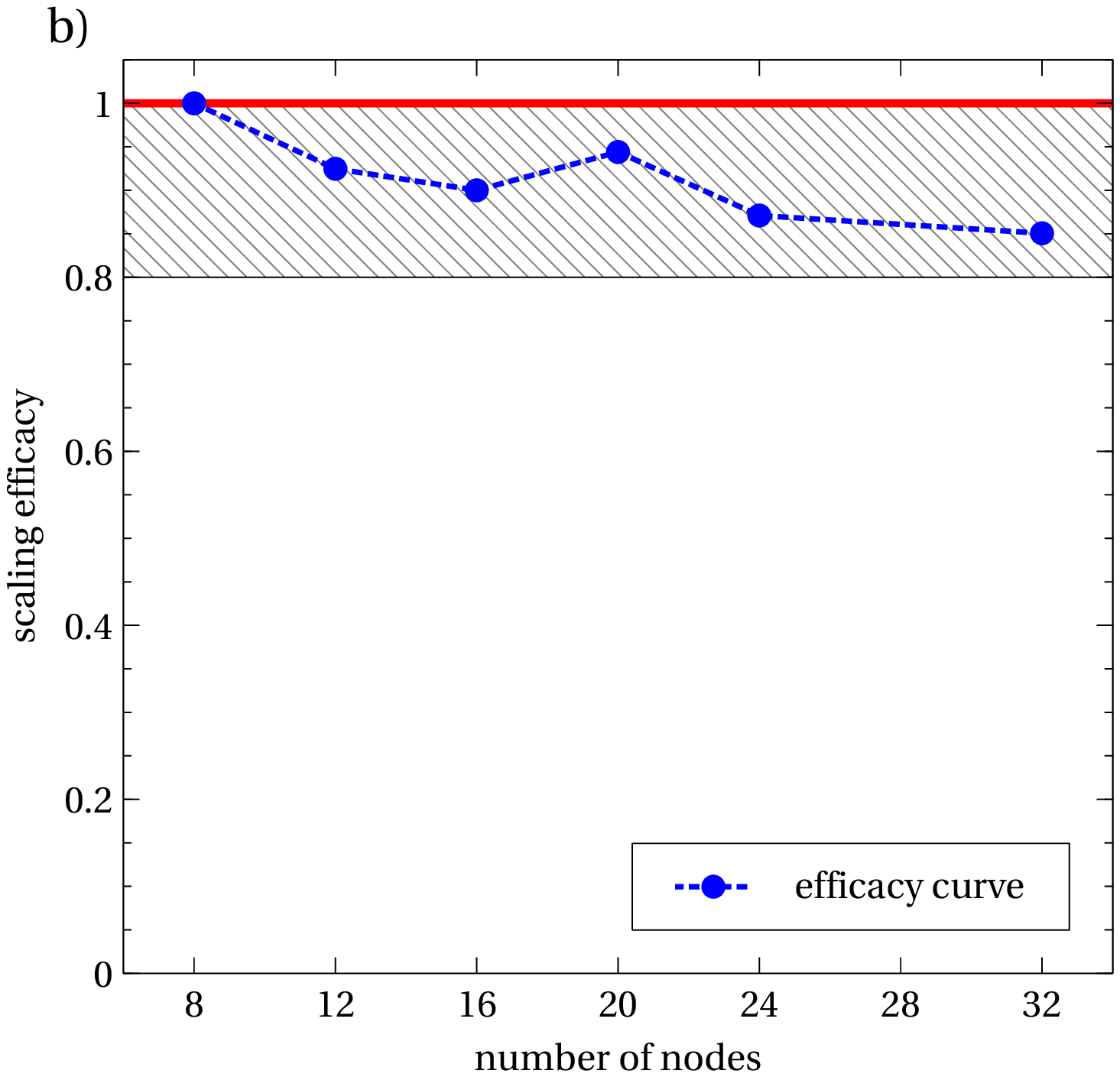}
\includegraphics[width=5.6cm]{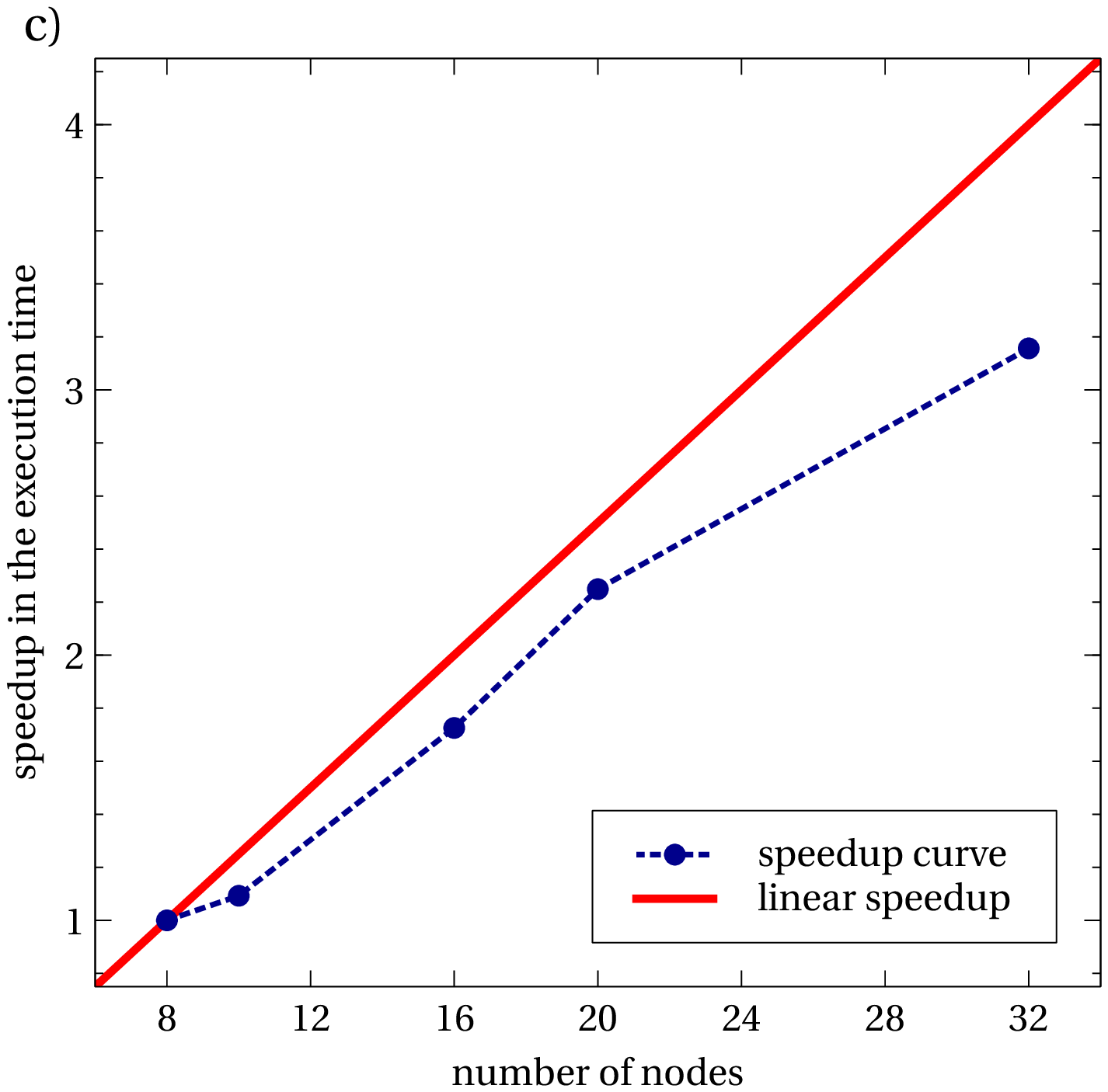}\hspace*{5mm}
\includegraphics[width=5.6cm]{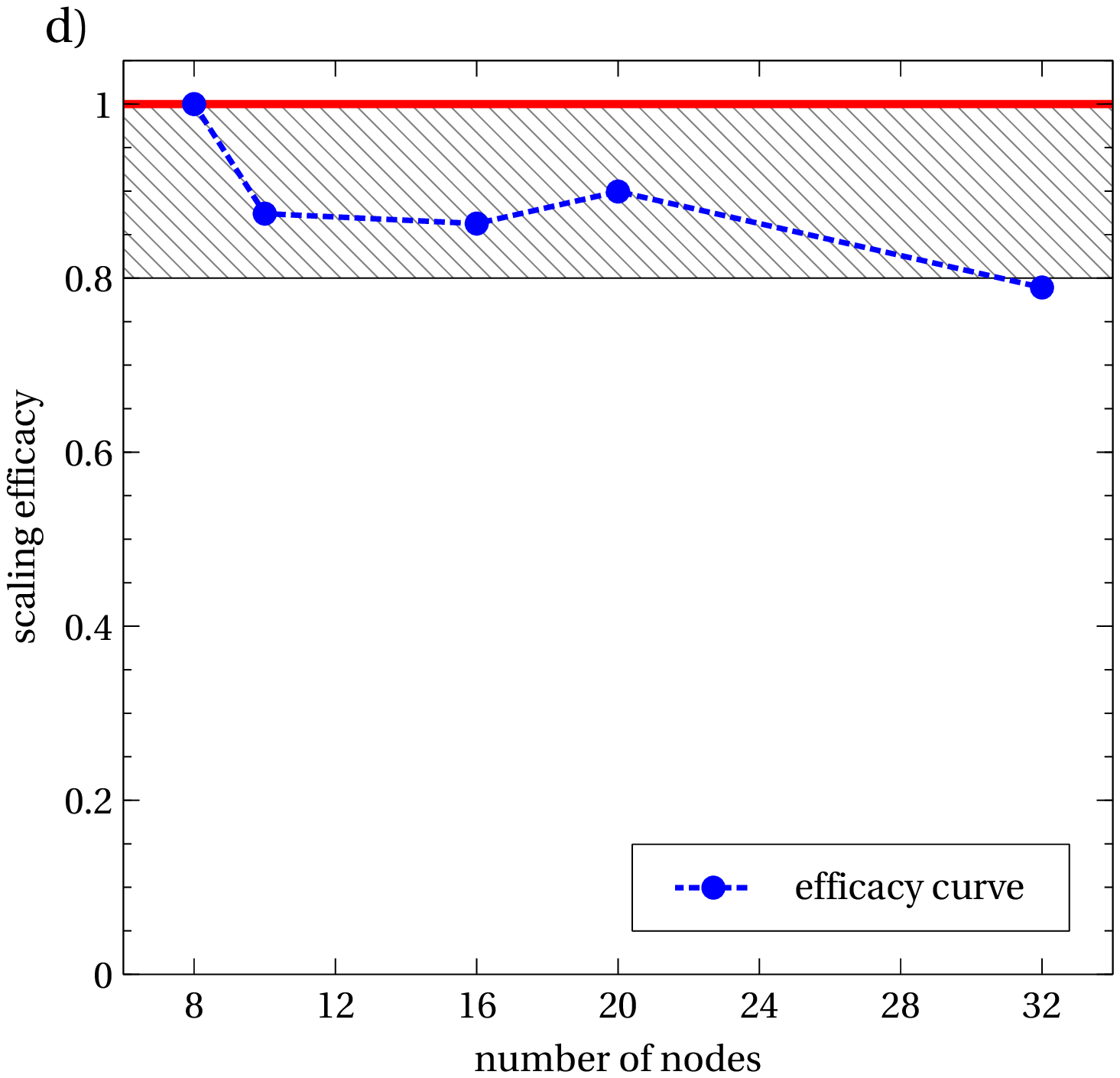}
\caption{Speedup in the execution time and efficacy curves of imagtime3d-hyb and realtime3d-hyb programs as a function of the number of MPI nodes used for large grid sizes. The results are obtained on a cluster with nodes containing 2 x 8-core Sandy Bridge Xeon 2.6 GHz processors with 32 GB of RAM and Infiniband QDR interconnect: (a) speedup of imagtime3d-hyb on a 1920x1600x1280 grid; (b) efficacy of imagtime3d-hyb on a 1920x1600x1280 grid; (c) speedup of realtime3d-hyb on a 1600x1280x960 grid; (d) efficacy of realtime3d-hyb on a 1600x1280x960 grid. Shaded areas in graphs (b) and (d) represent high-efficacy regions, where speedup is at least 80\% of the ideal one.}
\label{fig4}
\end{center}
\end{figure}

Figures~\ref{fig3} and \ref{fig4} show the scalability results obtained for hybrid versions of programs for small and large grid sizes as a function of number of MPI nodes used. The baseline for calculation of speedups in the execution time for small grid sizes are previous, purely OpenMP programs, while for large grid sizes, which cannot fit onto a single node, the baseline are hybrid programs with minimal configuration runs on 8 nodes. The figures also show efficacies, defined as percentages of measured speedups compared to the ideal ones. We see that an excellent scalability (larger than 80\% compared to the ideal one) can be obtained for up to 32 nodes. The tests have been performed on a cluster with nodes containing 2 x 8-core Sandy Bridge Xeon 2.6 GHz processors with 32 GB of RAM and Infiniband QDR (Quad Data Rate, 40 Gbps) interconnect. We stress that the scalability depends greatly on the ratio between the calculation and communication time per iteration, and has to be studied for a particular type of processors and interconnect technology.

\noindent\\
{\em Additional comments:} This package consists of 2 programs, see Program title above. Both are hybrid, threaded and distributed (OpenMP/MPI parallelized). For the 
particular purpose of each program, see descriptions below.

\noindent\\
{\em Running time:} All running times given in descriptions below refer to programs compiled with OpenMPI/GCC compiler and executed on 8 to 32 nodes with 2 x 8-core Sandy Bridge Xeon 2.6 GHz processors with 32 GB of RAM and Infiniband QDR interconnect. With the supplied input files for small grid sizes, running wallclock times of several minutes are required on 8 to 10 MPI nodes.

\noindent\\
{\em Special features:} (1) Since the condensate wavefunction data are distributed among the MPI nodes, when writing wavefunction output files each MPI process saves its data into a separate file, to avoid I/O issues. Concatenating the corresponding files from all MPI processes will created the complete wavefunction file. (2) Due to a known bug in OpenMPI up to version 1.8.4, allocation of memory for indexed datatype on a single node for large grids (such as 800x640x480) may fail. The fix for this bug is already in 3c489ea branch and is fixed in OpenMPI as of version 1.8.5. 

\noindent\\
Program summary (i)

\noindent\\
{\em Program title:} imagtime3d-hyb.\\
{\em Title of electronic files:} imagtime3d-hyb.c, imagtime3d-hyb.h.\\
{\em Computer:} Any modern computer with C language, OpenMP- and MPI-capable compiler installed.\\
{\em RAM memory requirements:} 300 MBytes of RAM for a small grid size 240x200x160, and scales with Nx*Ny*Nz. This is total amount of memory needed, and is distributed over MPI nodes used for execution.\\
{\em Programming language used:} C/OpenMP/MPI.\\
{\em Typical running time:} Few minutes with the supplied input files for a small grid size 240x200x160 on 8 nodes. Up to one hour for a large grid size 1920x1600x1280 on 32 nodes (1000 iterations).\\
{\em Nature of physical problem:} This program is designed to solve the time-dependent GP nonlinear partial differential equation in three space dimensions with an anisotropic trap. The GP equation describes the properties of a dilute trapped Bose-Einstein condensate.\\
{\em Method of solution:} The time-dependent GP equation is solved by the split-step Crank-Nicolson method by discretizing in space and time. The discretized equation is then solved by propagation in 
imaginary time over small time steps. The method yields solutions of stationary problems.\\

\noindent\\
Program summary (ii)

\noindent\\
{\em Program title:} realtime3d-hyb.\\
{\em Title of electronic files:} realtime3d-hyb.c, realtime3d-hyb.h.\\
{\em Computer:} Any modern computer with C language, OpenMP- and MPI-capable compiler installed.\\
{\em RAM memory requirements:} 410 MBytes of RAM for a small grid size 200x160x120, and scales with Nx*Ny*Nz. This is total amount of memory needed, and is distributed over MPI nodes used for execution.\\
{\em Programming language used:} C/OpenMP/MPI.\\
{\em Typical running time:}  10-15 minutes with the supplied input files for a small grid size 200x160x120 on 10 nodes. Up to one hour for a large grid size 1600x1280x960 on 32 nodes (1000 iterations).\\
{\em Nature of physical problem:} This program is designed to solve the time-dependent GP nonlinear partial differential equation in three space dimensions with an anisotropic trap. The GP equation describes the properties of a dilute trapped Bose-Einstein condensate.\\
{\em Method of solution:} The time-dependent GP equation is solved by the split-step Crank-Nicolson method by discretizing in space and time. The discretized equation is then solved by propagation in real time over small time steps. The method yields solutions of stationary and non-stationary problems.\\

\section*{Acknowledgements}
\noindent
 B.~S., V.~S., A.~B., and A.~B. acknowledge support by the 
Ministry of Education, Science, and Technological Development of the Republic of Serbia under 
projects ON171017, III43007, ON171009, ON174027 and IBEC, and by DAAD - German Academic and 
Exchange Service under project IBEC.
P.~M. acknowledges support by the Science and Engineering Research Board, Department of Science and Technology, Government of India under project No.~EMR/2014/000644.
S.~K.~A. acknowledges support by the CNPq of Brazil under project 303280/2014-0, and by the FAPESP of Brazil under project 2012/00451-0.
Numerical simulations were run on the PARADOX supercomputing facility
at the Scientific Computing Laboratory of the Institute of Physics
Belgrade, supported in part by the Ministry of Education, Science,
and Technological Development of the Republic of Serbia under project
ON171017.

\end{small}


\begin{thebibliography}{17}

\bibitem{C}
D. Vudragovi\'{c}, I. Vidanovi\'{c}, A. Bala\v{z}\corref{cor1}, P. Muruganandam, S.~K. Adhikari, C programs for solving the time-dependent Gross-Pitaevskii equation in a fully anisotropic trap,
Comput. Phys. Commun. {\bf 183} (2012) 2021.

\bibitem{Fortran}
P. Muruganandam and S.~K. Adhikari, Fortran programs for the time-dependent Gross-Pitaevskii equation in a fully anisotropic trap,
Comput. Phys. Commun. {\bf 180} (2009) 1888.

\bibitem{all}
R.~K. Kumar and P. Muruganandam, J. Phys. B: At. Mol. Opt. Phys. {\bf 45} (2012) 215301;\\
L.~E. Young-S. and S.~K. Adhikari, Phys. Rev. A {\bf 86} (2012) 063611;\\
S.~K. Adhikari, J. Phys. B: At. Mol. Opt. Phys. {\bf 45} (2012) 235303;\\
I. Vidanovi\'{c}, N.~J. van Druten, and M. Haque, New J. Phys. {\bf 15} (2013) 035008;\\
S. Balasubramanian, R. Ramaswamy, and A.~I. Nicolin, Rom. Rep. Phys. {\bf 65} (2013) 820;\\
L.~E. Young-S. and S.~K. Adhikari, Phys. Rev. A {\bf 87} (2013) 013618;\\
H. Al-Jibbouri, I. Vidanovic, A. Balaz, and A. Pelster, J. Phys. B: At. Mol. Opt. Phys. {\bf 46} (2013) 065303;\\
X. Antoine, W. Bao, and C. Besse, Comput. Phys. Commun. {\bf 184} (2013) 2621;\\
B. Nikoli\'{c}, A. Bala\v{z}, and A. Pelster, Phys. Rev. A {\bf 88} (2013) 013624;\\
H. Al-Jibbouri and A. Pelster, Phys. Rev. A {\bf 88} (2013) 033621;\\
S.~K. Adhikari, Phys. Rev. A {\bf 88} (2013) 043603;\\
J.~B. Sudharsan, R. Radha, and P. Muruganandam, J. Phys. B: At. Mol. Opt. Phys. {\bf 46} (2013) 155302;\\
R.~R. Sakhel, A.~R. Sakhel, and H.~B. Ghassib, J. Low Temp. Phys. {\bf 173} (2013) 177;\\
E.~J.~M. Madarassy and V.~T. Toth, Comput. Phys. Commun. {\bf 184} (2013) 1339;\\
R.~K. Kumar, P. Muruganandam, and B.~A. Malomed, J. Phys. B: At. Mol. Opt. Phys. {\bf 46} (2013) 175302;\\
W. Bao, Q. Tang, and Z. Xu, J. Comput. Phys. {\bf 235} (2013) 423;\\
A.~I. Nicolin, Proc. Rom. Acad. Ser. A-Math. Phys. {\bf 14} (2013) 35;\\
R.~M. Caplan, Comput. Phys. Commun. {\bf 184} (2013) 1250;\\
S.~K. Adhikari, J. Phys. B: At. Mol. Opt. Phys. {\bf 46} (2013) 115301;\\
\v{Z}. Marojevi\'{c}, E. G\"{o}kl\"{u}, and C. L\"{a}mmerzahl, Comput. Phys. Commun. {\bf 184} (2013) 1920;\\
X. Antoine and R. Duboscq, Comput. Phys. Commun. {\bf 185} (2014) 2969;\\
S.~K. Adhikari and L.~E. Young-S, J. Phys. B: At. Mol. Opt. Phys. {\bf 47} (2014) 015302;\\
K. Manikandan, P. Muruganandam, M. Senthilvelan, and M. Lakshmanan, Phys. Rev. E {\bf 90} (2014) 062905;\\
S.~K. Adhikari, Phys. Rev. A {\bf 90} (2014) 055601;\\
A. Bala\v{z}, R. Paun, A. I. Nicolin, S. Balasubramanian, and R. Ramaswamy, Phys. Rev. A {\bf 89} (2014) 023609;\\
S.~K. Adhikari, Phys. Rev. A {\bf 89} (2014) 013630;\\
J. Luo, Commun. Nonlinear Sci. Numer. Simul. {\bf 19} (2014) 3591;\\
S.~K. Adhikari, Phys. Rev. A {\bf 89} (2014) 043609;\\
K.-T. Xi, J. Li, and D.-N. Shi, Physica B {\bf 436} (2014) 149;\\
M.~C. Raportaru, J. Jovanovski, B. Jakimovski, D. Jakimovski, and A. Mishev, Rom. J. Phys. {\bf 59} (2014) 677;\\
S. Gautam and S.~K. Adhikari, Phys. Rev. A {\bf 90} (2014) 043619;\\
A.~I. Nicolin, A. Bala\v{z}, J. B. Sudharsan, and R. Radha, Rom. J. Phys. {\bf 59} (2014) 204;\\
K. Sakkaravarthi, T. Kanna, M. Vijayajayanthi, and M. Lakshmanan, Phys. Rev. E {\bf 90} (2014) 052912;\\
S.~K. Adhikari, J. Phys. B: At. Mol. Opt. Phys. {\bf 47} (2014) 225304;\\
R.~K. Kumar and P. Muruganandam, Numerical studies on vortices in rotating dipolar Bose-Einstein condensates, Proceedings of the 22nd International Laser Physics Workshop, J. Phys. Conf. Ser. {\bf 497} (2014) 012036;\\
A.~I. Nicolin and I. Rata, Density waves in dipolar Bose-Einstein condensates by means of symbolic computations, High-Performance Computing Infrastructure for South East Europe's Research Communities: Results of the HP-SEE User Forum 2012, in Springer Series: Modeling and Optimization in Science and Technologies {\bf 2} (2014) 15;\\
S.~K. Adhikari, Phys. Rev. A {\bf 89} (2014) 043615;\\
R.~K. Kumar and P. Muruganandam, Eur. Phys. J. D {\bf 68} (2014) 289;\\
J.~B. Sudharsan, R. Radha, H. Fabrelli, A. Gammal, and B.~A. Malomed, Phys. Rev. A {\bf 92} (2015) 053601;\\
S.~K. Adhikari, J. Phys. B: At. Mol. Opt. Phys. {\bf 48} (2015) 165303;\\
F.~I. Moxley III, T. Byrnes, B. Ma, Y. Yan, and W. Dai, J. Comput. Phys. {\bf 282} (2015) 303;\\
S.~K. Adhikari, Phys. Rev. E {\bf 92} (2015) 042926;\\
R.~R. Sakhel, A.~R. Sakhel, and H.~B. Ghassib, Physica B {\bf 478} (2015) 68;\\
S. Gautam and S.~K. Adhikari, Phys. Rev. A {\bf 92} (2015) 023616;\\
D. Novoa, D. Tommasini, and J.~A. N\'{o}voa-L\'{o}pez, Phys. Rev. E {\bf 91} (2015) 012904;\\
S. Gautam and S.~K. Adhikari, Laser Phys. Lett. {\bf 12} (2015) 045501;\\
K.-T. Xi, J. Li, and D.-N. Shi, Physica B {\bf 459} (2015) 6;\\
R.~K. Kumar, L.~E. Young-S., D. Vudragovi\'{c}, A. Bala\v{z}, P. Muruganandam, and S.~K. Adhikari, Comput. Phys. Commun. {\bf 195} (2015) 117;\\
S. Gautam and S.~K. Adhikari, Phys. Rev. A {\bf 91} (2015) 013624;\\
A. I. Nicolin, M.~C. Raportaru, and A. Bala\v{z}, Rom. Rep. Phys. {\bf 67} (2015) 143;\\
S. Gautam and S.~K. Adhikari, Phys. Rev. A {\bf 91} (2015) 063617;\\
E.~J.~M. Madarassy and V.~T. Toth, Phys. Rev. D {\bf 91} (2015) 044041.

\bibitem{OpenMPI}
Open Message Passing Interface (OpenMPI), \url{http://www.open-mpi.org/} (2015).

\bibitem{MPICH}
Message Passing Interface Chameleon (MPICH), \url{https://www.mpich.org/} (2015).

\bibitem{dongarra}
J. Choi, J.~J. Dongarra, D.~W. Walker,
Parallel matrix transpose algorithms on distributed memory concurrent computers,
Parallel Comput.  {\bf 21} (1995) 1387.
 

\end{thebibliography}
\end{document}